\documentclass[12pt,a4paper]{article}
\usepackage[utf8]{inputenc}
\usepackage[english]{babel}
\usepackage{amsmath}
\usepackage{amsfonts}
\usepackage{amssymb}
\usepackage{makeidx}
\usepackage{graphicx}
\usepackage[left=2cm,right=2cm,top=2cm,bottom=2cm]{geometry}
\usepackage{amsthm}
\newtheorem*{remark}{Remark}
\title{Covariant Brackets for Particles and Fields}
\date{}
\begin{document}
\maketitle
\author{\textbf{M. Asorey} \\ \textit{Departamento de F{\'{\i}}sica Te\'orica, Facultad de Ciencias, Universidad de Zaragoza}\\50009 Zaragoza, Spain}\\

\author{\textbf{F. M. Ciaglia and F. Di Cosmo} \\\textit{Dipartimento di Fisica “E. Pancini” dell Universitá “Federico II” di Napoli}\\
Complesso Universitario di Monte S. Angelo, via Cintia, 80126 Naples, Italy.\\
\textit{Sezione INFN di Napoli}\\
Complesso Universitario di Monte S. Angelo, via Cintia,80126 Naples, Italy}\\

\author{\textbf{A. Ibort} \\\textit{Departamento de Matem\'aticas, Universidad Carlos III de Madrid}\\
Avda. de la Universidad 30, 28911 Legan\'es, Madrid, Spain.\\
\textit{ICMAT, Instituto de Ciencias Matem\'aticas (CSIC - UAM - UC3M - UCM)}\\
Nicol\'as Cabrera,13–15, Campus de Cantoblanco, UAM, 28049, Madrid, Spain}

\begin{abstract}
A geometrical approach to the covariant formulation of the dynamics of relativistic  systems is introduced. A realization of Peierls brackets by means of a bivector field over the space of solutions of the Euler-Lagrange equations of a variational principle is presented. The method is illustrated with some relevant examples.

\end{abstract}

\section*{Introduction}
Since the advent of relativity the quest for a covariant bracket, which does not require a splitting of spacetime into space and time to be defined, has become a relevant issue in order to get a covariant physical description of fields and particles. 
In particular such a search is necessary  to understand how to formulate a quantum field theory which must be able to capture all the features of the corresponding classical theory.

The first steps in this direction were proposed by Peierls in 1952\cite{Peierls}. His pioneering work was devoted to the search of  a Poisson bracket which does not require either a canonical formalism or a splitting of spacetime. His method applies to  fields living on a suitable space-time manifold. In this framework particle mechanics is treated as a field theory in $0+1$-dimensions  over a one-dimensional temporal line.

A deeper analysis into the problems of a covariant formulation of field theories was carried out by DeWitt. In order to extend the Peierls bracket to the realm of gauge theories he carefully analyzed the consequences of the presence of symmetry groups  for the physical system. Furthermore a link between Peierls bracket and the effects of measurement processes was proposed also in classical mechanics (for a detailed exposition see the book by DeWitt \cite{DeWitt}).

Although the formulation of the idea  is clear, its geometrical interpretation can now be better understood at the light of new geometrical formulations of quantum theories\cite{Grabowski}. The aim of   this letter is to show the role of geometrical elements in the covariant formalism of particles and fields. 
We will  argue later that such a geometrical interpretation  is very useful  for unveiling the ambiguities of the formalism and to better understand its limits and generalizations. 
We will illustrate the formalism by means of  examples; a more detailed analysis will be carried out in future works. 

\section{Covariant Brackets}
\paragraph{Action Functional.}

 Let us consider a single particle in  classical  mechanics. From a kinematical point of view our {\sl fields} are the trajectories of a particle in a certain configuration space. In the simplest situation our configuration space is a vector space, for instance $\mathbb{R}^3 \times \mathbb{R} = \mathbb{R}^4$, where we add also the time dimension to be closer to the covariant formulation. A generic trajectory is a differentiable section $\gamma \, : \, \mathbb{R} \, \mapsto \, \mathbb{R}^4 \times \mathbb{R}$. So far we are not concerned with a possible non-trivial topology or differential structure of the space.

The fundamental point in this formulation is the choice of the action functional $S\left[ \gamma \right]$. It plays a double role: first,  it provides us with a set of equations of motion by means of the corresponding variational principle, and on the other hand it allows to define tangent vectors to the space of solutions of the equations of motion. Let us consider the action functional
of a particle with only a kinetic term, i.e.
\begin{equation}
S[\gamma] = \dfrac{m}{2} \int_{\mathbb{R}} g(\dot{\gamma}, \dot{\gamma}) ds \, ,
\label{Lagrangian for the free motion}
\end{equation}      
where $g$ is the background Riemannian metric of the configuration space, $m$ is the mass of the particle, and $\dot{\gamma}$ denotes the differential of the map $\gamma$. If we introduce a set of globally defined coordinate functions $\left\lbrace x^{\mu} \right\rbrace$ on $\mathbb{R}^4$, the functional \eqref{Lagrangian for the free motion} assumes the more familiar form
\begin{equation}
S[\gamma] = \dfrac{m}{2} \int_{\mathbb{R}} g_{\mu \nu} \left( x^{\mu} \right) \dfrac{dx^{\mu}}{ds} \dfrac{dx^{\nu}}{ds} ds \,. 
\label{Lagrangian for the free motion in coordinate}
\end{equation}     

According to the variational principle, the dynamical trajectories are stationary points of the action functional. The variation of the trajectory by means of a tangent vector field, $\delta \gamma$, gives rise  to the equations of  motion, 
\begin{equation}
\delta S = \int_{\mathbb{R}}\dfrac{d}{d\lambda}\mathcal{L}\left( x^{\mu}+\lambda \delta x^{\mu}, \dot{x}^{\mu} + \lambda \dfrac{d}{ds} \delta x^{\mu} \right) ds = 0 \,,
\end{equation}
where also the perturbation of the velocities is the differential of the perturbation along the path. 

For a complete set of variations\footnote{By completeness we are requiring the variations to be able to separate trajectories by the values of functionals.}, we get the  Euler-Lagrange equations 
\begin{equation}
\dfrac{d}{ds} \left( \dfrac{\partial \mathcal{L}}{\partial \dot{x^{\mu}}} \right) -\dfrac{\partial \mathcal{L}}{\partial x^{\mu}} = 0 \, , 
\label{Euler Lagrange equation}
\end{equation}      
which in the case of purely kinematic Lagrangian density become the geodesics equation
\begin{equation}
\dfrac{d^2x^{\mu}}{ds^2}+\Gamma_{\nu \rho}^{\mu}\dfrac{d{x}^{\nu}}{ds}\dfrac{d{x}^{\rho}}{ds} = 0\,,
\label{Euler Lagrange equation Free motion}
\end{equation}
where $\Gamma_{\nu \rho}^{\mu}$ are the Christoffel symbols of the Levi-Civita connection associated with the metric tensor $g=g_{\mu \nu}dx^{\mu}\otimes dx^{\nu}$.
\begin{remark}
Actually we should consider this perturbation in a homotopy class of paths, but  we will deal with these technical details in a future work. In general a variational principle can be settled by fixing a fiducial path $\gamma_0$ and considering another path in the same homotopy class of the referring one. Since these paths enclose an area, we can define a functional directly in terms of a two form, and provide a variational principle by looking for its stationary points. In other words we are looking for critical values of fluxes. This generalization is very useful when dealing with systems which do not admit an intrinsic Lagrangian formulation, e.g. the motion of an electron in the magnetic field generated by a monopole \cite{unfolding}$^-$\cite{qb} .
\end{remark}
Let us now choose a solution $\gamma_0$ of Euler-Lagrange equations which will be our reference point in the space of trajectories. We can use the action functional to define a tangent space to this solution. Indeed, we can consider the set of functionals which are defined on this solution; by means of one of these functionals, e.g. $A$, we build up a new variational principle in terms of the modified action $S' = S + \lambda A $. The new Euler-Lagrange equations are written in a compact form as:
\begin{equation}
\left( \delta S + \lambda \delta A \right)\left[ \gamma_A \right] = 0 \,.
\end{equation}
If we are interested in the description of small perturbations $\delta_A \gamma$ with respect to the reference solution, the Euler-Lagrange equations simplify and we get the  DeWitt 
 equations of the small perturbations, i.e.
\begin{equation}
\left.\left( C_{\mu \nu}\dfrac{d^2}{ds^2} + D_{\mu \nu}\dfrac{d}{ds} + E_{\mu \nu} \right)\right|_{\gamma=\gamma_0}\delta_Ax^{\nu} = - \left. \left[ \dfrac{\partial \mathcal{A}}{\partial x^{\mu}} -\dfrac{d}{ds}\left( \dfrac{\partial \mathcal{A}}{\partial \dot{x}^{\mu}} \right)\right] \right|_{\gamma = \gamma_0} \, ,
\label{Equation of small perturbations}
\end{equation} 
where we assumed that  $A$ is given in terms of a Lagrangian density $\mathcal{A}$, i.e. $A=\int \mathcal{A}\, ds $.
The operator on the l.h.s of Eq.\eqref{Equation of small perturbations} is written in terms of the following matrices
\begin{eqnarray}
& C_{\mu \nu} = - \dfrac{\partial^2 \mathcal{L}}{\partial \dot{x}^{\mu}\partial \dot{x}^{\nu}} \\
& D_{\mu \nu} = -\dfrac{d}{ds}\dfrac{\partial^2 \mathcal{L}}{\partial \dot{x}^{\mu}\partial \dot{x}^{\nu}} - \dfrac{\partial^2 \mathcal{L}}{\partial \dot{x}^{\mu}\partial x^{\nu}} + \dfrac{\partial^2 \mathcal{L}}{\partial x^{\mu}\partial \dot{x}^{\nu}} \\
& E_{\mu \nu} = - \dfrac{d}{ds}\dfrac{\partial^2 \mathcal{L}}{\partial \dot{x}^{\mu}\partial x^{\nu}} + \dfrac{\partial^2 \mathcal{L}}{\partial x^{\mu}\partial x^{\nu}}
\end{eqnarray}      
evaluated along the reference solution $\gamma_0$. 

Applying these formulas to the case of free motion we obtain the following equations
\begin{equation}
\dfrac{\nabla^2}{ds^2}\delta_A x^{\nu} + R(\delta_A \gamma, \dot{\gamma}_0)_{\mu}^{\nu}\dot{x}_0^{\mu} = g^{\mu \nu}\left. \left[ \dfrac{\partial \mathcal{A}}{\partial x^{\mu}} -\dfrac{d}{ds}\left( \dfrac{\partial \mathcal{A}}{\partial \dot{x}^{\mu}} \right)\right] \right|_{\gamma = \gamma_0} \, ,
\label{Jacobi equation}
\end{equation}
where the symbol $\frac{\nabla^2}{dt^2}$ denotes the second covariant derivative along the vector field $\dot{\gamma}_0$, and $R$ is the Riemann curvature, both associated with the Levi-Civita connection.
On the l.h.s. we immediately recognize the operator $\mathcal{J}$ which appears in Jacobi equation \cite{Bautista}. 
It is important to stress that these equations allow to define variations which are associated with suitable functionals. The existence of zero-modes of the Jacobi equation points out the existence of nodal points.

\paragraph{Peierls Brackets: Definition and Geometrical Interpretation.}
Peierls idea is to proceed by selecting two different solutions of Eq. \eqref{Jacobi equation}, one vanishing at the far past along the trajectory, the other  one vanishing in the far future, and then take their difference. The result is a solution of Jacobi equation with no source. This tangent vector may act as a linear operator on the other functionals. The result of this action is called Peierls bracket. If  $\delta_A^- x^{\mu}$ and $\delta_A^+ x^{\mu}$ denote the required solutions, the Peierls bracket of two functionals $A, \, B$  is
\begin{equation}
\left\lbrace A , B \right\rbrace_{P} = \int_{\mathbb{R}} \left( \delta_A^+ x^{\mu} -\delta^-_A x^{\mu}\right)\left. \left[ \dfrac{\partial \mathcal{B}}{\partial x^{\mu}} -\dfrac{d}{ds}\left( \dfrac{\partial \mathcal{B}}{\partial \dot{x}^{\mu}} \right)\right] \right|_{\gamma = \gamma_0} ds\,.
\label{eqn: peierls bracket}
\end{equation}

The solutions $\delta_A^\pm x^{\mu} $ matching the asymptotic past/future conditions can be expressed in terms of the retarded/advanced  Green functions 
\begin{eqnarray}
G_\pm^{\nu \mu}(s,s')=\left(L_{\mu\nu}^{\pm}\right)^{-1}(s,s'),
\end{eqnarray}
where
\begin{eqnarray}
L_{\mu\nu}=\left.\left( C_{\mu \nu}\dfrac{d^2}{ds^2} + D_{\mu \nu}\dfrac{d}{ds} + E_{\mu \nu} \right)\right|_{\gamma=\gamma_0}.
\end{eqnarray}
 Indeed, in that case
\begin{eqnarray}
&\delta_A^+x^{\nu}(s) = \int_{\mathbb{R}}G_+^{\nu \mu}(s,s')\left[ \dfrac{\partial \mathcal{A}}{\partial x^{\mu}} -\dfrac{d}{ds}\left( \dfrac{\partial \mathcal{A}}{\partial \dot{x}^{\mu}} \right) \right] (s')ds'\\ 
&\delta_A^-x^{\nu}(s) = \int_{\mathbb{R}}G_-^{\nu \mu}(s,s')\left[\dfrac{\partial \mathcal{A}}{\partial x^{\mu}} -\dfrac{d}{ds}\left( \dfrac{\partial \mathcal{A}}{\partial \dot{x}^{\mu}} \right)\right] (s')ds'\,.
\end{eqnarray}

One may wonder why we are choosing the difference $ \delta_A^+ x^{\mu} - \delta_A^- x^{\mu}$  of these solutions   in order to define Peierls brackets. The explanation is simple. Let us consider the following difference: 
$$
 g_{\mu \nu}J^{\mu}_1 \left( \dfrac{\nabla^2}{ds^2}J_2^{\nu}+ R(J_2, \dot{\gamma}_0)_{\rho}^{\nu}\dot{x}_0^{\rho} \right) - g_{\mu \nu}J^{\mu}_2 \left( \dfrac{\nabla^2}{ds^2}J_1^{\nu}+ R(J_1, \dot{\gamma}_0)_{\nu}^{\rho}\dot{x}_0^{\nu} \right) = 
$$
\begin{equation}
 = \dfrac{\nabla}{ds} \left( g_{\mu \nu}J_1^{\mu}\dfrac{\nabla}{ds}J_2^{\nu}- g_{\mu \nu}J_2^{\mu}\dfrac{\nabla}{ds}J_1^{\nu} \right) = \dfrac{d}{ds}\left( \omega_{\mathcal{L}}\left( \left(J_1,\dfrac{d}{ds}J_1\right),\left(J_2,\dfrac{d}{ds}J_2\right) \right) \right)
\label{Symplectic structure}
\end{equation} 
where $\omega_{L}$ in the final expression represents a two-form on the tangent bundle $T\mathbb{R}^4$ associated with the Lagrangian density $\mathcal{L}$ which is given explicitly by:
$$
\omega_{\mathcal{L}}= \dfrac{\partial^2 \mathcal{L}}{\partial \dot{x}^{\mu}\partial \dot{x}^{\nu}} dx^{\mu}\wedge dv^{\nu} + \left( \dfrac{\partial^2 \mathcal{L}}{\partial \dot{x}^{\mu}\partial x^{\nu}} - \dfrac{\partial^2 \mathcal{L}}{\partial x^{\mu}\partial \dot{x}^{\nu}} \right) dx^{\mu}\wedge dx^{\nu}\,.
$$

If we now put $J_1=\delta_A^+ x^{\mu} - \delta_A^- x^{\mu}$ and $J_2=\delta_B^+ x^{\mu} - \delta_B^- x^{\mu}$, where $\delta_B^+ x^{\mu} - \delta_B^- x^{\mu}$ is the solution associated with a second functional $B$, we obtain 
\begin{equation}
g_{\mu \nu}J^{\mu}_1 \left( \dfrac{\nabla^2}{ds^2}J_2^{\nu}+ R(J_2, \dot{\gamma}_0)_{\rho}^{\nu}\dot{x}_0^{\rho} \right) - g_{\mu \nu}J^{\mu}_2 \left( \dfrac{\nabla^2}{ds^2}J_1^{\nu}+ R(J_1, \dot{\gamma}_0)_{\nu}^{\rho}\dot{x}_0^{\nu} \right) = 0 \,.
\end{equation}
Consequently 
\begin{equation}
\dfrac{d}{ds}\left(\omega_{\mathcal{L}}\left( \left(J_1,\dfrac{d}{ds}J_1\right),\left(J_2,\dfrac{d}{ds}J_2\right) \right) \right)=0\, ,
\label{Conserved Bracket}
\end{equation}
which means that the quantity $\textstyle\omega_{\mathcal{L}}( (J_1,\textstyle\frac{d}{ds}J_1),(J_2,\textstyle\frac{d}{ds}J_2) )$ is preserved along solutions of the referring equations of the motion.

This result actually expresses a general fact: Eq.\eqref{Conserved Bracket} remains valid for every action functional coming from a Lagrangian density not only of the kinetic type. Essentially, it is related to the symmetry of the second variation of the action, and when we have a Lagrangian density it gives rise to the two-form $\omega_{\mathcal{L}}$. It could be interesting to analyze what kind of conservation law would have been obtained for more general path functionals. However the answer to this question is postponed to future work.

We will show now that the preserved quantity does coincide with Peierls bracket. Indeed, Peierls bracket can be written as:
\begin{equation}
\left\lbrace A,B \right\rbrace_P = \int_{\mathbb{R}}\delta_A^-x^j\left( \dfrac{\partial \mathcal{B}}{\partial x^j}-\dfrac{d}{ds} \left( \dfrac{\partial \mathcal{B}}{\partial \dot{x}^j} \right) \right)ds - \int_{\mathbb{R}}\delta_B^-x^j\left( \dfrac{\partial \mathcal{A}}{\partial x^j}-\dfrac{d}{ds} \left( \dfrac{\partial \mathcal{A}}{\partial \dot{x}^j} \right) \right)ds\, .
\end{equation}
This definition and the one in Eq. \eqref{eqn: peierls bracket} coincide when the action of the variation $\delta_A^+ x^{\mu}$ over the functional $B$ equals the action of the variation $\delta_B^- x^{\mu}$ over the functional $A$ \cite{DeWitt}.

Replacing  $J_1=\delta_A^- x^{\mu}=G^-(\delta A)$ \footnote{This means that we are selecting a particular solution of the perturbed equation, which is given in terms of  the advanced Green function.} and $J_2=\delta_B^- x^{\mu} = G^-(\delta B)$ in Eq.\eqref{Symplectic structure}, we get 
\begin{equation}
\left\langle \delta B , G^-(\delta A) \right\rangle (s) - \left\langle \delta A , G^-(\delta B) \right\rangle(s) = \dfrac{d}{ds}\left( \omega_{\mathcal{L}}\left( G^-(\delta A), G^-(\delta B) \right) \right) \,.
\end{equation}
Integrating both sides of the equation along the reference path, we get 
\begin{equation}
\left\lbrace A,B \right\rbrace_P = \lim_{s \rightarrow -\infty} \omega_{\mathcal{L}}\left( G^-(\delta A), G^-(\delta B) \right)(s) \,,
\end{equation}
which does coincide with the previous conserved quantity because we are in the remote past, where the effect of the solution $\delta_A^+ x^{\mu}$ is negligible. Since these expressions coincide we can actually compute the conserved quantity at any point of the trajectory using the variations written in terms of the difference $\delta_A^+ x^{\mu} - \delta_A^- x^{\mu}$ which we denote by $\tilde{G}(\delta A)$.

Let us now look at the homogeneous part of the linearized problem \eqref{Equation of small perturbations}. Since it is a linear differential equation, the space of its solutions is a vector space. As it is a second order ordinary differential equation, the dimension of this vector space is $2\times 4=8$, where $4$ is the dimension of the configuration space. We can find a basis of this space by looking for a set of $8$ independent solutions. Indeed, when the equations 
\begin{eqnarray}
&L_{\mu \nu}J^{\nu} = 0\\
&J^{\mu}(s_1)=J^{\mu}(s_2)=0
\end{eqnarray}   
have  only a trivial solution, we can find a set of independent solutions by solving the eight homogeneous differential equations
\begin{eqnarray}
&L_{\mu \nu}J_-^{\nu} = 0\\
&J^{(\rho)}_-(-T)=J^{\rho}_-\\
&J^{(\rho)}_-(T)=0
\label{Independent Solutions 1}
\end{eqnarray}   
and
\begin{eqnarray}
&L_{\mu \nu}J_+^{\nu} = 0\\
&J^{(\rho)}_+(T)=J^{\rho}_+\\
&J^{(\rho)}_+(-T)=0 \,
\label{Indipendent Solutions 2}
\end{eqnarray}   
where the vectors $J_+^{\rho}$ and $J_-^{\rho}$ have only one non-vanishing component  in the $\rho$-position with value $1$. 

In terms of these solutions the Green's function $\tilde{G}^{\mu \nu}(s,s')$ can be written as 
\begin{equation}
\tilde{G}^{\mu \nu}(s,s') = \sum_{\rho}\dfrac{J_+^{(\rho)\mu}(s)J_-^{(\rho)\nu}(s')-J_-^{(\rho)\mu}(s)J_+^{(\rho)\nu}(s')}{W(J^{(\rho)}_+,J^{(\rho)}_-)}
\end{equation}  
where $W(J^{(\rho)}_+,J^{(\rho)}_-)= \omega_{\mathcal{L}}\left( J^{(\rho)}_+,J^{(\rho)}_- \right)$ that we already know is a constant of the motion. 

This choice of variations is actually a choice of a basis of the tangent space to the reference solution $\gamma_0$, seen as a solution of a second order differential equation. We will denote this basis by $\left\lbrace \frac{\partial}{\partial x_+^{\rho}} , \frac{\partial}{\partial x_-^{\rho}} \right\rbrace$, where $x_+$ and $x_-$ are the parameters labelling the particular solution.

If we now replace this expression in the definition of $\delta_A^+ x^{\mu} - \delta_A^- x^{\mu}$ we obtain:
\begin{equation}
\nonumber
\delta_A^+x^{\mu} - \delta_A^-x^{\mu} = \sum_{\rho}\dfrac{J_+^{(\rho)\mu}(s)\int_{\mathbb{R}}J_-^{(\rho)\nu}(s')\frac{\delta A}{\delta x^{\nu}}(s')ds'-J_-^{(\rho)\mu}(s)\int_{\mathbb{R}}J_+^{(\rho)\nu}(s')\frac{\delta A}{\delta x^{\nu}}(s')ds'}{W(J^{(\rho)}_+,J^{(\rho)}_-)} \, ,
\end{equation} 
and its action over a functional $B$ can be written as
\begin{eqnarray}
\nonumber
\left\lbrace A , B \right\rbrace_P& = \displaystyle \sum_{\rho} \dfrac{1}{W(J^{(\rho)}_+,J^{(\rho)}_-)}\left( \int_{\mathbb{R}}\dfrac{\delta B}{\delta x^{\mu}}(s) J_+^{(\rho)\mu}(s) ds \right) \left( \int_{\mathbb{R}}J_-^{(\rho)\nu}(s')\dfrac{\delta A}{\delta x^{\nu}}(s')ds'\right) \\
\nonumber
& -  \sum_{\rho} \dfrac{1}{W(J^{(\rho)}_+,J^{(\rho)}_-)} \left( \int_{\mathbb{R}}\dfrac{\delta B}{\delta x^{\mu}}(s) J_-^{(\rho)\mu}(s) ds \right) \left( \int_{\mathbb{R}}J_+^{(\rho)\nu}(s')\dfrac{\delta A}{\delta x^{\nu}}(s')ds'\right) \,.
\end{eqnarray}

We can identify variations along a path with the restriction of a vector field on the space of solutions, so that this bilinear operation on functionals can be represented in terms of the bivector field along the space of solutions 
\begin{equation}
\Lambda = \sum_{\rho} \dfrac{1}{W(J^{(\rho)}_+,J^{(\rho)}_-)}\dfrac{\partial}{\partial x_+^{\rho}} \wedge \dfrac{\partial}{\partial x_-^{\rho}}\,.
\end{equation}
From this expression we show that  Peierls bracket is a bilinear antisymmetric operation. However, this does not guarantees that it defines  a Poisson structure. 

In summary, we have shown that Peierls bracket can be read actually as a bivector field defined on the space of solutions of a system of differential equations in the case of single particle mechanics. Moreover if the dynamics is covariant with respect to the action of the Poincar\'e group, this action is directly implemented as a Hamiltonian symmetry of the system because it maps solutions into solutions.  

Let us now analyze some examples that can be useful for clarifying the meaning of Peierls brackets. 

Let us start with the geodesic motion in a flat Euclidean space, $\mathbb{R}^4$. The Euler-Lagrange equations are 
\begin{equation}
\delta_{jk}\dfrac{d^2}{ds^2}x^{k}=0
\end{equation}
and the operator of the Jacobi equation Eq.\eqref{Jacobi equation} reduces to 
\begin{equation}
\mathcal{J} = \delta_{lk}\dfrac{d^2}{ds^2}J^k \,.
\end{equation}  

Let us consider now a path which is solution of the variational problem and which passes through two established points:
\begin{equation}
x^j(s)= \dfrac{x^j_-+x^j_+}{2}+\dfrac{x^j_--x^j_+}{2T}s.
\end{equation}
The particular integration constants  have been chosen  to match the conditions 
\begin{eqnarray}
&x(-T)=x_-\\
&x(T)=x_+\,.
\label{Boundary Condiion free particle}
\end{eqnarray}

As previously illustrated we have to add a source term to the Jacobi equation \eqref{Jacobi equation} to find two particular solutions, one vanishing in the far past and one in the far future. These two solutions are respectively obtained by convolution with the advanced and retarded Green's functions given by
\begin{eqnarray}
&G_+^{jk}(s-s')=\delta^{jk}\,\theta(s-s') \, (s-s') \\
&G_-^{jk}(s-s')=\delta^{jk}\,\theta(s'-s) \, (s'-s)\,.
\end{eqnarray}
In the definition of Peierls brackets we have the difference of Green functions $\Delta^{ij}=G^{ij}_+-G^{ij}_-$ which defines the causal Green's function \cite{Sorkin}
$$
\Delta^{ij}(s,s')=\delta^{ij} \,(s-s').
$$
If we consider a perturbation given in terms of a density function $\mathcal{A}$ we obtain that the required difference is
\begin{equation}
\left(\delta_A^+x^j-\delta_A^-x^j\right) (s) =\delta^{jk}\int_{\mathbb{R}}(s-s')\left( \dfrac{\partial\mathcal{A}}{\partial x^k}-\dfrac{d}{ds'}\dfrac{\partial\mathcal{A}}{\partial \dot{x}^k} \right)(s') ds'\,.
\end{equation}
The Peierls bracket of two functionals given by  two densities $\mathcal{A}$ and $\mathcal{B}$ is thus 
\begin{equation}
\nonumber
\left\lbrace A , B \right\rbrace_P =\delta^{jk}\int_{\mathbb{R}} \int_{\mathbb{R}}(s-s')\left( \dfrac{\partial\mathcal{B}}{\partial x^j}-\dfrac{d}{ds}\dfrac{\partial\mathcal{B}}{\partial \dot{x}^j} \right)(s)\left( \dfrac{\partial\mathcal{A}}{\partial x^k}-\dfrac{d}{ds'}\dfrac{\partial\mathcal{A}}{\partial \dot{x}^k} \right)(s') ds ds'\,,
\end{equation}
and an explicit calculation shows that this expression actually does coincide with the conserved quantity
\begin{equation}
\omega_{\mathcal{L}}\left( \left(J_1,\dfrac{d}{ds}J_1\right),\left(J_2,\dfrac{d}{ds}J_2\right) \right)
\end{equation}
when $J_1=\delta_A^+x^j-\delta_A^-x^j$ and $J_2 = \delta_B^+x^j-\delta_B^-x^j$.

Let us now choose a set of independent solutions of the systems \eqref{Independent Solutions 1} and \eqref{Indipendent Solutions 2}. A straightforward analysis 
gives
\begin{eqnarray}
&J_+^{(j)}(s)=x^{(j)}_+(s+T) \\
&J_-^{(j)}(s)=x^{(j)}_-(s-T) 
\end{eqnarray} 
where
\begin{eqnarray}
(x^{(j)}_\pm)^i=\begin{cases}
{1} & {\hbox{if }\    i=j}\\
{0}& {\hbox{if  }\  i\neq j}.
\end{cases}
\end{eqnarray}
Since $\omega_{\mathcal{L}}(J_+^{(j)},J_-^{(j)}) = 2T$ we get that the bivector field defining the Peierls bracket is  
\begin{equation}
\Lambda = \sum_j T \dfrac{\partial}{\partial x_+^j}\wedge \dfrac{\partial }{\partial x_-^j}\,.
\end{equation} 

We can immediately recognize that the tensor $\Lambda$ defines a Poisson bracket: it does coincide 
with the push-forward of the Poisson tensor $\Lambda_{\mathcal{L}}=\omega_{\mathcal{L}}^{-1}$ with respect to the canonical transformation defined by 
the Hamilton principal function
\begin{equation}
S(x_-,x_+) = \sum_{j}\dfrac{(x_+^j-x_-^j)^2}{2T}\,
\end{equation} 
associated with the Lagrangian density generating the equations of motion \cite{Lanczos}.

\section{Field Theory: Peierls brackets for Non Relativistic Quantum Mechanics}
Before addressing the case of a relativistic covariant theory, we consider the case of a non-relativistic field theory: non-relativistic quantum mechanics.
When physical states are identified with vectors of a Hilbert space there are two different descriptions of the theory: either in terms of  one-dimensional trajectories in a Hilbert space \cite{Asorey}, or as a field theory in terms of  complex valued fields over a four dimensional spacetime.
In the first case we  have a one dimensional space of parameters,  $\mathbb{R}$ and  our fields are sections $\psi \, : \, \mathbb{R} \, \mapsto \, \mathbb{R}\times \mathbb{H}$ where $\mathbb{H}$ is the Hilbert space of quantum states. 

\paragraph{Hilbert space representation.}
Pure States in Quantum Mechanics are actually rays of this Hilbert space, that is points in the Projective Hilbert space. However, since the scope of this example is to introduce the action of the unitary group it is not essential to consider this extra constraint and we shall consider a description in terms of vectors of a Hilbert space.

The action functional is
\begin{equation}
S[\psi] = \int_{\mathbb{R}} \left[ i \left( \Big\langle \psi \Big |  \frac{d\psi}{ds}\Big\rangle -\Big\langle \frac{d\psi}{ds} \Big | \psi \Big\rangle \right) - \langle \psi | H | \psi \rangle \right] ds \,, 
\label{Action Functional Quantum Mechanics}
\end{equation}   
where $H$ is a self-adjoint operator on the Hilbert space $\mathbb{H}$.  Euler-Lagrange equations become the Schr\"{o}dinger equations:
\begin{equation}
2i \dfrac{d }{ds}|\psi \rangle = H | \psi \rangle
\label{Schrodinger Equation}
\end{equation}  
and
\begin{equation}
2i \dfrac{d }{ds}\langle \psi | = \langle \psi | H \,.
\label{Adjoint Schrodinger Equation}
\end{equation}  
There is a set of constraints which imply that these equations are not independent; in fact one is the adjoint of the other. However since we will consider only unitary actions these constraints are automatically preserved and we can consider $|\psi\rangle$ and $\langle \psi |$ as conjugate variables. A more careful description of constrained dynamics will be presented elsewhere. 

Let us introduce a basis $|e_n \rangle$ of the Hilbert space $\mathbb{H}$ labelled by an integer $n$. The vector $|\psi \rangle=\sum_{n=0}^\infty\psi^n |e_n\rangle$ is expressed in terms of its components $\psi^n$ and we can introduce the new set of coordinates:
\begin{eqnarray}
&q^n = \mathfrak{Re} \, \psi^n \\
&p_n = \mathfrak{Im} \, \psi^n
\end{eqnarray}
The Lagrangian can now be rewritten as 
\begin{equation}
\mathcal{L} = q^n \dot{p}_n - p_n \dot{q}^n + q_k H_j^k q^j + p_k H_j^k p^j \, ,
\end{equation} 
where we are using Einstein's summation convention.
For the moment let us forget about the Hamiltonian operator. 
Eventually our Lagrangian is just
$$
\mathcal{L} = q^n \dot{p}_n - p_n \dot{q}^n
$$
and the associated Euler-Lagrange equations are
\begin{equation}
\delta_{jk}\dot{p}^k = 0  \qquad \qquad -\delta_{jk}\dot{q}^k  = 0
\end{equation}
Solutions are parametrized in terms of an initial condition $\left( q_0^k,p^k_0 \right)$. Let us now consider a perturbation given in terms of the quadratic form of a self-adjoint operator, that is 
$$
A = \int_{\mathbb{R}} \left(q_k A_j^k q^j + p_k A_j^k p^j \right) ds\,.
$$
The equation of the small perturbations becomes
\begin{equation}
\left(
\begin{array}{cc}
0 & \mathbb{I} \\
-\mathbb{I} & 0 
\end{array}
\right)
\dfrac{d}{dt} \left( 
\begin{array}{c}
\delta_A q\\
\delta_A p
\end{array}
\right) = -
\left(
\begin{array}{cc}
A & 0 \\
0 & A 
\end{array}
\right)
\left(
\begin{array}{c}
q_{0}\\
p_{0}
\end{array}
\right) \,,
\end{equation}
and the associated Green function $\tilde{G}$ is the following constant matrix:
\begin{equation}
\tilde{G} =
\left(  
\begin{array}{cc}
0 & -\mathbb{I} \\
\mathbb{I} & 0
\end{array}
\right)\,.
\end{equation}

From this expression we get the following solution for the vector field associated with the functional $A$:
\begin{equation}
\delta_A^+ \psi - \delta_A^-\psi = A^{jk}q_{0j}\dfrac{\partial}{\partial p_0^k} - A^{jk}p_{0j}\dfrac{\partial }{\partial q_0^k}\,.
\end{equation}
Therefore  the Peierls bracket of  two functionals $A,\,B$, whose densities are quadratic forms, becomes
\begin{equation}
\left\lbrace A, B \right\rbrace_P = -p_0^t BA q_0 + q_0^t BA p_0 = p_0^t (AB - BA)q_0 = p_0^t\, [A,B]\, q_0
\label{Peierls bracket for Hilbert space QM}
\end{equation}
where $q^t_0$ ($p^{t}_{0}$) denotes the transpose of the vector whose components are $q_0^j$ ($p^{j}_{0}$).   

Unlike the case of classical particle mechanics, the space of solutions is in one-to-one correspondence with the whole Hilbert space and the bracket  is  defined for any pair of functions over an infinite dimensional vector space. However a consequence of our choice of the Lagrangian is that Peierls bracket does coincide with the canonical Poisson bracket over this separable Hilbert space, which is
\begin{equation}
\Lambda = \delta^{jk}\dfrac{\partial}{\partial p_0^j} \wedge \dfrac{\partial}{\partial q_0^k}\,.
\end{equation} 
Indeed the Lagrangian function $\mathcal{L}=p\dot{q}-q\dot{p}$ is a ``canonical'' Lagrangian function on the tangent bundle $T\mathbb{M}$ of a symplectic vector space $\mathbb{M}$ and in this case Peierls bracket will always coincide with the canonical Poisson structure over the same space.

\paragraph{Position Representation.}
Let us now come back to our initial action functional \eqref{Action Functional Quantum Mechanics} and  realize the abstract Hilbert space $\mathbb{H}$ as the space of square-integrable functions over a three-dimensional euclidean space $\mathbb{R}^3$. We will identify the parameter $t$ as another component of our spacetime which now becomes $\mathbb{R}^4$. Vectors of the Hilbert space are eventually realized as functions $\psi ({\bf x},t)$. If we consider the Hamiltonian operator
\begin{equation}
H=-\sum_{j=1}^3 \dfrac{\partial^2}{\partial x^{j2}} 
\end{equation}  
associated with the free motion, the Lagrangian in \eqref{Action Functional Quantum Mechanics} becomes
\begin{equation}
\mathcal{L} = i\left( \psi^*\dfrac{\partial \psi}{\partial t} - \dfrac{\partial \psi^*}{\partial t} \psi \right)(\mathbf{x},t) + \sum_{j=1}^3 \dfrac{\partial \psi^*}{\partial x^j}\dfrac{\partial \psi}{\partial x^j}(\mathbf{x},t)\,.
\end{equation}

Therefore, in position representation, the  theory which was previously defined over a 1-dimensional parameter space, is realized as a field theory over a four dimensional spacetime. However it is important to remark that in non relativistic quantum mechanics space and time coordinates play two  completely different roles: the former are represented as (unbounded) multiplication operators over the Hilbert space and the parameters appearing in the wave function are the eigenvalues of these operators; the latter is only a parameter which labels the group of transformations defining the dynamics.   

A solution of this equation can be obtained by considering an initial wave function $\psi(\mathbf{x}, t_0)$ and the action  of the  one parameter dynamical group of unitary transformations $U(t-t_0)$ generated by $H$, that is:
\begin{equation}
|\psi(t)\rangle = U(t-t_0)|\psi(t_0)\rangle
\end{equation} 
If we consider the position representation the wave function at the instant $t$ is written in terms of the one at $t_0$ according to the following formula:
\begin{equation}
\psi({\bf x}, t) = \int_{\mathbb{R}^3}G(x,x_0;t-t_0)\psi({\bf x},t_0) dx'
\end{equation} 
where the kernel $G(x,x_0;t-t_0)$ in the integral is the Green function associated with the evolution operator $U(t-t_0)$. For the free motion we already know the expression of this operator\cite{Esposito}
\begin{equation}
G(x-x',t-t_0) = -\sqrt{i}\left( \dfrac{1}{2\pi (t-t_0)} \right)^{\frac{3}{2}} \exp{\left( i\frac{\sum_j(x^j-x_0^j)^2}{2(t-t_0)} \right)}\,.
\end{equation}
Let us now consider a functional $A[\psi]$ whose density is a quadratic form associated with a time-independent self-adjoint operator $\mathbf{A}$, that is 
\begin{equation}
A[\psi]= \int_{\mathbb{R}^4\times \mathbb{R}^4}\psi^*({\bf x}',t')\, \mathcal{A}(x,x')\, \psi({\bf x},t)\, d{\bf x}\,dt\,d{\bf x}'\,dt' \,,
\end{equation}   
where $\mathcal{A}(x,x')$ is the expression of $\mathbf{A}$ in the position representation.
We will compute the variation associated with the functional $A$ noticing that in this case we can identify variations with vectors of the Hilbert space $\mathbb{H}$. In general, the commutator Green function $\tilde{G}(t,t_0)$ for the abstract Schr\"{o}dinger equation is $-iU(t-t_0)$ and consequently its expression in the position representation is the Green kernel $-i G(x-x_0;t-t_0)$. The variation $\delta_A\psi$ is  a solution of small perturbations equation with source given by $-\mathbf{A}\, U(t-t_0)\, \psi(t_0)$, whereas its adjoint is the source in the equation for $\delta_A\psi^*$. 
We thus have 
\begin{equation}
\delta_A\psi(t) = i \int_\mathbb{R}U(t-\tau)\mathbf{A}U(\tau-t_0)\psi(t_0) d\tau
\end{equation}    
and the adjoint equation for $\delta_A\psi^*(t)$. 
The Peierls bracket of the functionals $A, B$ associated with the operators $\mathbf{A},\mathbf{B}$ is given by:
\begin{equation}
\left\lbrace A,B \right\rbrace_P = -i\int_{\mathbb{R}}dt\int_{\mathbb{R}}d\tau \langle \psi(t_0)|\left[ \mathbf{A}_H(\tau-t_0) , \mathbf{B}_H(t-t_0) \right] |\psi(t_0) \rangle
\end{equation}
where $\mathbf{A}_H(\tau) = U^{\dagger}(\tau)\,\mathbf{A}\,U(\tau)$ denotes the Heisenberg evolution of the operator $\mathbf{A}$, and simillarly for $\mathbf{B}_H(t)$. Because of the invariance under time translation we can replace $t_0$ with $0$. Then in the position representation we can write the following result
\begin{equation}
\left\lbrace A, B \right\rbrace_P = \mathfrak{Im}  \int_{\mathbb{R}^4}d{\bf x}dt \int_{\mathbb{R}^4}d{\bf y}d\tau \int_{\mathbb{R}^3}d{\bf \xi} \psi^*(y,0)\mathbf{B}_{H}(y,\xi,t)\mathbf{A}_{H}(\xi,x,\tau)\psi(x,0)  \,.
\end{equation}

\paragraph{Geometrical Interpretation.} 
Applying  equation \eqref{Symplectic structure} to this case, we obtain the following conservation law \cite{AIS}:
\begin{equation}
\dfrac{d}{ds}\left( \langle J_1(s) | J_2(s) \rangle - \langle J_2(s) | J_1(s) \rangle \right) = 0 \,,
\end{equation} 
and we immediately notice that the expression in brackets is the canonical symplectic structure on a Hilbert space. If we realize this Hilbert space as the space $\mathbb{H}=\mathcal{L}^2\left( \mathbb{R}^3,d{\bf x} \right)$ of square-integrable functions on $\mathbb{R}^3$ with respect to Lebesgue measure $d{\bf x}$, the conserved object is written as
\begin{equation}
\omega \left( J_1 , J_2 \right) = i \int_{\mathbb{R}^3}\left( J_1^*({\bf x} , t)J_2({\bf x} , t) - J_2^*({\bf x} , t)J_1({\bf x} , t) \right) d{\bf x}\,.
\end{equation} 
Let us now choose $J_1 = \delta_A^+\psi -\delta_A^-\psi = \tilde{G}(\delta A)$ and $J_2 = \delta_B^+\psi -\delta_B^-\psi = \tilde{G}(\delta B)$. Since $\tilde{G}(t-t_0)=-iU(t-t_0)$ we can write the Green function as the superposition of eigenfunctions of the Hamiltonian operator. Indeed, for the Hamiltonian operator associated with the free motion, we have that:
\begin{equation}
G(x-x_1,t-\tau) = \dfrac{1}{L^3(2\pi)^{\frac{3}{2}}}\sum_{{\bf k}} e^{ -ik_j(x^j-k^{j}t)}e^{ ik_j(x_1^j-k^{j}\tau)} \,,
\end{equation}
where we have considered our system to be confined in a cubic box $\Omega$ of edge $L$ with periodic boundary conditions in order to avoid the introduction of the projection-valued operator measure over the spectrum of the Hamiltonian operator. 
Since 
\begin{equation}
\int_{\Omega} e^{-i(k_j-k'_j)x^j} d{\bf x} = \delta_{k,k'}
\end{equation}
we get once more that the Peierls bracket can be written in terms of the bivector
\begin{equation}
\lambda = i \sum_{{\bf k}}\dfrac{\partial }{\partial \psi^*_{{\bf k}}}\wedge \dfrac{\partial}{\partial \psi_{{\bf k}}}
\end{equation}
where the components $\psi_{{\bf k}}$ are computed over a basis of forward propagating modes, whereas the complex conjugate is meant for backward propagating modes. The result is determined by the choice of the commutator Green's function, which allows us to construct variations which are tangent to the space of the solutions. 

The sharp splitting in spatial and temporal degrees of freedom is a consequence of the fact that the theory is non relativistic. We will briefly explain how to proceed in the relativistic case. 

Indeed, for a field theory the conservation rule \eqref{Symplectic structure} will be expressed as the vanishing of the divergence of a current. If the spacetime admits a time function defining simultaneity surfaces, the integration of the current on a simultaneity surface is preserved and it provides us with a symplectic structure over the space of solutions of the equations of motion. The use of the commutator Green function $\tilde{G}$ allows us to write the Peierls bracket in terms of forward and backward propagating modes. However in the following paragraph we will look at a concrete example: the relativistic theory for a scalar real field. 

\paragraph{Relativistic Scalar Field Theory.}
Let us now consider  a relativistic real scalar field $\phi$ over a four dimensional Minkowski spacetime $\left( \mathbb{R}^4, \eta \right)$ with free action functional
\begin{equation}
S[\phi]=\int_{\mathbb{R}^4}\dfrac{1}{2}\left( \eta^{\mu \nu}\partial_{\mu}\phi \partial_{\nu}\phi - m^2 \phi^2 \right) d^4x\,.
\end{equation} 
The Euler Lagrange equation  is the  Klein-Gordon equation:
\begin{equation}
\partial_{\mu}\partial^{\mu}\phi + m^2\phi = 0\,.
\end{equation}
If we consider the coordinate function $x_0$ as a globally defined time function $\tau$ the level sets of which define simultaneity surfaces, a solution of Klein-Gordon equation whose values at $\tau =\tau_1$ and $\tau = \tau_2$ are given by the two functions $\phi_1({\bf x})$ and $\phi_2({\bf x})$ respectively, is
\begin{equation} 
\nonumber
\phi({\bf x},\tau) = \dfrac{1}{(2\pi)^3}\int_{\mathbb{R}^3}\dfrac{d{\bf k\   e^{i{\bf k}{\bf x}}}}{\sin \omega_k (\tau_2 - \tau_1)} \left( \tilde{\phi}_1({\bf k})\sin \omega_{k}(\tau-\tau_2) - \tilde{\phi}_2({\bf k})\sin \omega_{k}(\tau-\tau_1) \right)\,.
\end{equation}
This can be seen as the superposition of harmonic oscillators with $k$-dependent frequencies. 
The conserved current \eqref{Symplectic structure} in this case is obtained as follows:
\begin{equation}
J_2 \left( \partial_{\mu}\partial^{\mu}J_1 \right) - \left( \partial_{\mu}\partial^{\mu}J_2 \right) J_1 = \partial^{\mu}\left( J_2\left( \partial_{\mu}J_1\right) - \left( \partial_{\mu}J_2 \right) J_1\right)\,, 
\end{equation}
and if we integrate the right hand side over the volume contained between two simultaneity surfaces we get that the flux of the current over one simultaneity surface is actually preserved along the solution. The Peierls bracket of two functionals $A$ and $B$ is defined  replacing $J_1$ and $J_2$ by $\tilde{G}(\delta A)$ and $\tilde{G}(\delta B)$ respectively. In this case we can write the commutator Green function as
\begin{equation}
\tilde{G}(x,y) = \int_{\mathbb{R}^3}\dfrac{d{\bf k}}{(2\pi)^3\omega_k}e^{i{\bf k}({\bf x}-{\bf y})}\left( \sin(\omega_k x_0)\cos(\omega_k y_0) - \cos(\omega_k x_0)\sin(\omega_ky_0) \right)\,.
\end{equation}
After some lenghty computations we can write the Peierls bracket  as 
\begin{equation}
\left\lbrace A , B \right\rbrace_P =\int_{\mathbb{R}^4}d^4z \int_{\mathbb{R}^4}d^4y\int_{\mathbb{R}^3}\dfrac{d{\bf k}}{(2\pi)^3\omega_k}e^{i{\bf k}({\bf y}-{\bf z})} \sin (\omega_k(y_0-z_0)) \dfrac{\delta A}{\delta \phi}(y)\dfrac{\delta B}{\delta \phi}(z)\,,
\end{equation} 
which can be seen as the superposition over the modes labelled by ${\bf k}$ of Peierls brackets for an harmonic oscillator. Furthermore each one of these modes can be treated as the superposition of a forward propagating mode and a backward propagating mode, $J_+$ and $J_-$ respectively. Causality is respected as can be shown by computing the brackets between the two functionals $\phi(x_1) = \int_{\mathbb{R}^4}\phi(x)\delta(x-x_1)$ and $\phi(x_2) = \int_{\mathbb{R}^4}\phi(x)\delta(x-x_2)$, that is  \cite{Asoreyb}:
\begin{equation}
\left\lbrace \phi(x_1),\phi(x_2) \right\rbrace_P = \tilde{G}(x_1-x_2)\,.
\end{equation}

\section{Conclusions and Outlooks} 
We have presented a geometrical analysis of Peierls brackets. As we have shown, the fundamental ingredient is the choice of a background action functional allowing us to select a subset of the space of fields consisting of the critical points of the functional itself, and to define tangent vectors to these points. We have highlighted the principal ideas in the simple case of the dynamics of point particles in classical mechanics and free scalar field theories.

It is not obvious whether or not  the bivector field which is associated with Peierls bracket on the space of solutions defines a Poisson structure. Indeed it is not always true that the space of solutions of a differential equation admits a Poisson structure, e.g. the space of light-like geodesics possesses a Jacobi structure as illustrated in the work by Bautista et al.\cite{Bautista}. A detailed analysis of the necessary conditions ensuring that Peierls bracket defines a Poisson structure is under study. Some sufficient conditions have been given by Forger and Romero \cite{Forger} and Khavkine \cite{Khavkine}.
 
A second point is related to gauge theories. As we have seen, another fundamental ingredient in this covariant formulation is the correspondence between the space of solutions of a differential equation and the space of initial conditions. In presence of gauge symmetries the same initial conditions can be evolved according to different dynamical trajectories. However the relationship between Peierls bracket and the symplectic structure $\omega_{\mathcal{L}}$ on the space of solutions suggests a way of treating the problem as exposed in the work by Dubrovin et al\cite{Dubrovin}.

Finally we remark that, although we restricted ourselves to consider only functionals associated with local densities, the formulation outlined here does not rely on this assumption. It would be relevant to extend the procedure presented here to more general functionals. A first step in this direction may be the introduction of functionals which are defined as two forms.
This however requires the selection of a reference path as illustrated in the work of Zaccaria et al\cite{Balachandran}.       

\subsection*{Acknowledgements}
We thank to G. Marmo for suggesting the geometric analysis of Peierls brackets.  We also thank A. P. Balachandran, 
for many enlightening discussions.   The work of M. A. has been partially supported by the 
Spanish  MINECO/FEDER grant FPA2015-65745-P and DGA-FSE grant 2015-E24/2 and the COST Action  MP1405 
QSPACE, supported by COST (European Cooperation in Science and Technology).

\bibliographystyle{plain}
{}

\end{document}